\documentclass{tlp}

\usepackage[T1]{fontenc}
\usepackage{orcidlink}
\usepackage{pifont}
\usepackage{xspace}
\usepackage[frozencache,cachedir=minted-cache]{minted}

\usepackage{tikz}
\usetikzlibrary{calc,positioning,decorations.text}

\usepackage[l3]{csvsimple}
\usepackage{booktabs}
\usepackage{rotating} 
\usepackage{adjustbox}

\usepackage{stmaryrd}
\usepackage{siunitx}
\usepackage{xstring}
\newcommand{\fmt}[1]{%
  \IfStrEq{#1}{OOM}{OOM}{%
    \IfStrEq{#1}{TO}{TO}{%
      #1%
    }%
  }%
}

\newcommand{\mgu}{\mathit{mgu}}
\newcommand{\match}{\mathit{match}}
\newcommand{\power}{\mathcal{P}}
\newcommand{\fpower}{\mathcal{P}_f}
\newcommand{\varset}{\mathcal{V}}
\newcommand{\vars}{\mathit{vars}}

\newcommand{\Sharing}{\textsc{Sharing}\xspace}

\newcommand{\Lin}{\textsc{Lin}\xspace}
\newcommand{\Free}{\textsc{Free}\xspace}
\newcommand{\ShLin}{\textsc{ShLin}\xspace}

\newcommand{\ShFrLin}{\textsc{ShFrLin}\xspace}
\newcommand{\KingShLin}{\textsc{ShLin}\textsuperscript{2}\xspace}

\newcommand{\prolog}[1]{\texttt{\detokenize{#1}}}
\newcommand{\module}[1]{\texttt{\detokenize{#1}}}

\newcommand{\PLAI}{\textsf{PLAI}\xspace}
\newcommand{\CiaoPP}{\textsf{CiaoPP}\xspace}
\newcommand{\Ciao}{\textsf{Ciao}\xspace}
\newcommand{\SWI}{\textsf{SWI-Prolog}\xspace}
\newcommand{\PROLOG}{\textsc{Prolog}\xspace}

\newcommand{\shl}{\mathit{sl}}
\newcommand{\matchshl}{\mathsf{match}_\shl}

\newminted{prolog}{fontsize=\small}

\newcommand{\xmark}{{\color{red}{\ding{55}}}}
\newcommand{\checkmark}{{\color{olive}{\ding{51}}}}

\tikzset{
    substitution/.style={draw, rectangle, minimum width=3cm, fill=gray!30, text=blue},
    operation/.style={circle, draw, fill=green!30}
}

\begin{document}

\jnlPage{1}{xx}
\jnlDoiYr{2026}
\doival{10.1017/xxxxx}

\title[Experimental evaluation of optimal abstract operators for sharing and linearity]{Experimental evaluation of optimal abstract operators for sharing and linearity analysis}

\begin{authgrp}
  \author{\sn{Amato} \gn{Gianluca}
  \orcidlink{0000-0002-6214-5198}}
  \affiliation{University of Chieti--Pescara\\
  \email{gianluca.amato@unich.it}}
  \author{\sn{Scozzari} \gn{Francesca}
  \orcidlink{0000-0002-2105-4855}}
  \affiliation{University of Chieti--Pescara\\  
  \email{francesca.scozzari@unich.it}}
\end{authgrp}


\maketitle

\begin{abstract}
In the field of static analysis of logic programs, the optimality of abstract operators is a valuable theoretical property, as it provides insight into the structure of abstract domains and the maximum precision that can be achieved. However, implementing optimal operators is often complex and may significantly impact performance, giving rise to a trade-off between precision and efficiency.

We experimentally investigate this trade-off in the context of sharing and linearity analysis of logic programs. Our experiments build on previous work that proposed several optimal operators for unification and matching. We have implemented these abstract operators and the corresponding abstract domains within the PLAI analyzer, part of the CiaoPP preprocessor, and we report the impact of increasing operator precision on the accuracy and performance of the overall analysis.
\end{abstract}

\begin{keywords}
abstract interpretation, logic programming, sharing, linearity
\end{keywords}

\section{Introduction}

This paper is concerned with the static analysis of logic programs, in particular with \emph{(set) sharing} analysis, whose aim is to identify sets of variables that share a common variable in a substitution.
For example, consider the substitution $\{x/f(u, v), y/g(u, u, u), z/v\}$. We say that $x$ and $y$ share the variable $u$, while $x$ and $z$ share the variable $v$, and no single variable is shared by $x$, $y$, and $z$. Sharing has been extensively studied in many papers, such as those by \cite{HansW92-tr, JacobsL92-jlp,MuthukumarH92-jlp,CodishSS99-toplas, BagnaraHZ02-tcs, JurjioRivasMLH24-tplp}. It is well-known that the precision of sharing analysis may be increased if we also record information about the linearity of terms.
We say that a term is \emph{linear} if it does not contain multiple occurrences of the same variable. For example, the term $f(x, f(y, z))$ is linear, while $f(x, f(y, x))$ is not, since $x$ occurs twice.
Integration of linearity and sharing information has been proposed by many authors, such as \cite{CodishDY91-iclp,HansW92-tr,MuthukumarH92-jlp,King94-esop},
and recently also in object-oriented languages by \cite{AmatoMS22-mscs}. The works of \cite{BagnaraZH05-tplp} and \cite{CodishMBB+95-toplas} contain extensive comparative evaluations of static analyses of logic programs combining many different abstract properties.

In this paper, we focus on the following abstract domains:
\begin{itemize}
    \item \Sharing, the basic domain for (set) sharing analysis, including the improvements presented by \cite{AmatoS09-tplp}.
    \item \ShLin, a simple domain combining sharing and linearity information (more precisely, it is the reduced product $\Sharing \times \Lin$), including the optimal operators presented by \cite{AmatoS10-tplp,AmatoS2014-tplp,AmatoS24-tplp}.
    \item \KingShLin, originally introduced by \cite{King94-esop} and later refined by \cite{AmatoS10-tplp,AmatoS24-tplp}.
\end{itemize}
For each domain, we implemented both the standard (as described in the literature) and the optimal abstract operators. 
An operator is optimal if it is the most precise correct abstraction of its concrete counterpart. While optimality is well understood theoretically, its practical impact is less clear.
Intuitively, increasing precision comes at a computational cost, since it often requires more complex operations. Therefore, one may expect a trade-off between precision and performance when adopting optimal operators. However, in some cases, higher precision may lead to smaller abstract objects, which can reduce the overall cost of the analysis. 

%
%
%
All the domains have been implemented in the \PLAI static analyzer \citep{MuthukumarH92-jlp}, which is part of \CiaoPP, the pre-processor \citep{BuenoLMPH25-ciaopp} of the \Ciao Prolog system \citep{BuenoCHLM97-ciao}. \PLAI performs a goal-dependent analysis, where the program is analyzed with respect to a given goal.
Goal-dependent analysis requires the implementation of two main operations called \emph{forward} and \emph{backward} unification\footnote{\cite{Bruynooghe91-jlp} and \cite{HansW92-tr} use \emph{procedure entry} and \emph{procedure exit}. \cite{MuthukumarH91-iclp} use \texttt{call\_to\_entry} and \texttt{exit\_to\_success}.} 
(see the papers by \cite{CortesiFW96-jlp} and \cite{AmatoS09-tplp}).
Forward unification performs parameter passing by unifying the atom chosen by the selection rule and the \emph{call substitution} with the head of the chosen clause. The result is an initial substitution for the clause, called \emph{entry substitution}. Backward unification propagates back to the initial goal the \emph{exit substitution} (that is, the result of the analysis of the clause body), thus obtaining the \emph{answer substitution}\footnote{Also called \emph{success substitution}.} for the initial goal.
Forward and backward unification may be implemented using a common unification operator ($\mgu$)  together with other straightforward operations like projection and renaming of variables. \cite{AmatoS09-tplp} show that precision of backward unification may be improved by making use of an additional matching ($\match$) operator.


The aim of this paper is: (1) to analyze the trade-off between standard and optimal operators for $\mgu$ and $\match$; (2) to explore alternative implementations of backward unification, based either on $\mgu$ alone or on a combination of $\mgu$ and $\match$; and (3) to experimentally evaluate the \KingShLin domain, which has not been implemented before.

The rest of the paper is structured as follows. Section~\ref{sec:sharing-domains} introduces the abstract domains considered in this work. Section~\ref{sec:implementation} describes their implementation within the \PLAI analyzer. Section~\ref{sec:experimental-evaluation} presents the experimental evaluation and discusses the impact of different operators on precision and performance. Finally, Section~\ref{sec:conclusions} 
concludes with directions for future work.

\section{Preliminaries: Abstract Domains for Sharing Analysis}
\label{sec:sharing-domains}
In this section, we provide a formal yet intuitive presentation of the abstract domains involved in the experimental evaluation.
%
Given two sets $C$ and $A$ of concrete objects and abstract properties, called the \emph{concrete} and \emph{abstract} domain
respectively, an \emph{abstract interpretation} \citep{CousotC92-jlc} is
given by an approximation relation $\rightslice \subseteq A \times C$.
When $a \rightslice c$ holds, we say that $a$ is a \emph{correct
abstraction} of $c$. We work in a framework where $A$ is endowed with 
a partial order relation to compare properties by \emph{precision}: $a_1 < a_2$ means that $a_1$ is more precise than $a_2$. The precision and approximation relations are related by the following: (1) $a \rightslice c$ and $a \leq a'$ imply $a' \rightslice c$, (2) each $c$ has a least (most precise) correct abstraction in $A$.
%
Following the collecting semantics approach \citep{AmatoMS20-tcs}, abstract operators are defined as correct approximations of their concrete counterparts.
Given a function $f: C \rightarrow C$, we say that $\tilde{f}: A \rightarrow A$ is a correct 
abstraction of $f$, and we write $\tilde{f} \rightslice f$, when
$
   a \rightslice c  \Rightarrow \tilde{f}(a)
   \rightslice f(c)
$. 
We say that $\tilde{f}: A \rightarrow A$ is the \emph{optimal} abstraction of $f$ when 
it is correct and, for each $f': A \rightarrow A$,
$
   f' \rightslice f
  \Rightarrow \tilde{f} \leq f'
$ with the pointwise ordering. In other words, the optimal abstraction of a concrete operator $f$ is 
its most precise correct approximation.

\subsection{The \texorpdfstring{\textsl{\Sharing}}{Sharing} abstract domain}
\label{sec:sharing}
The pioneering abstract domain \Sharing by \cite{JacobsL92-jlp} represents sharing information in a concise set representation. Let $\varset$ be an infinite set of variables and $\fpower(\varset)$ be the set of its finite subsets. The domain \Sharing can be defined as follows:
\begin{equation*}
 \Sharing = \{ [A,U]\mid A\subseteq\power(U),
        (A\neq\emptyset \Rightarrow \emptyset \in A), 
        U \in \fpower(\varset) \} \enspace .
\end{equation*}
Intuitively, an abstract object $[A,U]$ describes the relationships among the variables in $U$: if $S\in A$, the variables in $S$ are allowed to share a common variable. For example, $[\{ \{x,y\},\{z\},\emptyset \},\{w,x,y,z\}]$ represents the substitutions where $x$ and $y$ may possibly share, while $z$ is independent from both $x$ and $y$.  Moreover, the variable $w$, which does not appear in the first component, is ground. Two examples of such substitutions are $\{x/y, w/a\}$ and $\{w/a\}$ while $\{x/z, w/a\}$ is not, because $x$ and $z$ do share.  In addition, a special value $\bot$ is needed to denote the empty set of substitutions. 

\subsection{The \texorpdfstring{\textsl{\ShLin}}{ShLin} abstract domain}
The abstract domain \ShLin keeps track of the amount of linearity information in a substitution, coupling an object of \Sharing with the set of linear variables. More formally, it is defined as the reduced product of \Sharing and a basic linearity domain \Lin. Each element of \ShLin is a triple where the first component represents the sets of variables which may share, exactly as in \Sharing; the second component records the set of variables that must be linear; the third component is, as before, the
set of variables of interest. 
The abstract domain \ShLin is defined as follows, where  $\vars(S)$ denotes the set of all the variables in $S$:
\begin{equation*}
  \ShLin = \{ [S,L,U]\mid S \subseteq\power(U), (S \neq\emptyset \Rightarrow
  \emptyset \in S), L \supseteq U \setminus \vars(S), U \in \fpower(\varset)
  \} \enspace .
\end{equation*}
For instance, $[\{ \{x,y\}, \emptyset \},\{x,z\}, \{x,y,z\} ]$ represents the substitutions where $x$ may share with $y$, while $x$ and $z$ are (always) bound to a linear term (actually $z$ must be ground). Two examples of such substitutions are $\{x/t(u,a), y/t(u,u), z/a\}$ and $\{x/y, z/a\}$, while $\{x/t(u,u), y/t(u,u), z/a\}$ is not since $t(u, u)$ is not linear.

\subsection{Intuition on linearity and its impact on sharing precision}
\label{sec:linearity}
Linearity information plays a key role in improving the precision of sharing analysis.
In fact, when a variable occurs multiple times in a term, unification may introduce additional  sharing between variables.

For instance, consider the substitution $\theta_1=\{x/f(u,v)\}$ over the set $U=\{x\}$ of variables of interest. This substitution is abstracted in \Sharing as 
$S=[\{ \{x\}, \emptyset \},U]$. If we now compute the $\mgu$ between $S$ and $\mathcal E=\{x=f(y,z)\}$, we obtain the abstract object 
$\mgu(S,\mathcal E)=[\{ \{x,y\},\{x,z\},\{x,y,z\}, \emptyset \},\{x,y,z\}]$, which contains the sharing group $\{x,y,z\}$, due to the fact that the substitution $\theta_2=\{x/f(u,u)\}$ is also abstracted as $S$ in \Sharing.

Linearity information mitigates this problem. Intuitively, linearity acts as a constraint that rules out combinations of sharing groups that would only arise due to repeated occurrences of variables. 
In fact, the substitutions $\theta_1$ and $\theta_2$ are abstracted in different objects in \ShLin, namely 
$S_1=[\{ \{x\}, \emptyset \},\{x\},U]$
and
$S_2=[\{ \{x\},\emptyset \},\emptyset,U]$. The abstract unification operator in \ShLin can now exploit this information and obtain 
$\mgu(S_1,\mathcal E)=[\{ \{x,y\},\{x,z\},\emptyset \},
\{x,y,z\},\{x,y,z\}]$ while 
$\mgu(S_2,\mathcal E)=[\{ \{x,y\},\{x,z\},\{x,y,z\},\emptyset \},
\emptyset,\{x,y,z\}]$ which still contains the sharing group 
$\{x,y,z\}$.



\subsection{The \texorpdfstring{\textsl{\KingShLin}}{ShLin²} abstract domain}
The idea behind \KingShLin is to enhance \Sharing by annotating each sharing group with linearity information on each variable. If a variable in a sharing group is (possibly) non-linear, it is denoted using $\omega$ as an exponent, otherwise the variable is linear.
For instance, the object $xy^\omega z$ (called a 2-sharing group) represents the sharing group $\{x,y,z\}$ and the information that the common variable shared among $x$, $y$ and $z$ may occur more than once in $y$ but only once in $x$ and $z$. For example, this is the case for the variable $u$ in the substitution $\{ x/f(u,v,v), y/g(u,u,w), z/u \}$.
%
The \KingShLin abstract domain\footnote{Usually the definition also requires that the set $S$ is downward closed w.r.t. linearity information: for instance, the set $\{xy^\omega z\}$ is not downward closed, while $\{xyz, xy^\omega z\}$ is downward closed. For details, refer to the work by \cite{AmatoS10-tplp}.} 
can be defined as follows:
\begin{equation*}
 \KingShLin = 
  \bigl\{ [S,U] \mid \text{$S$ is a set of 2-sharing groups},
  U \in \fpower(\varset), S \neq \emptyset \Rightarrow \emptyset \in S \bigr \}
  \enspace .
\end{equation*}
Linearity information may vary between 2-sharing groups. For instance, $[\{xy, y^\omega z, \emptyset \}, \{x, y, z\}]$ represents the substitutions $\theta$ where the variable $y$ may share with $x$ or $z$, $x$ and $z$ are linear, while $y$ might be non-linear. However, the variable in common between $\theta(x)$ and $\theta(y)$ may only appear once in $\theta(y)$. An example of such a substitution is $\{x/u, y/t(u,v,v), z/v\}$, while $\{x/u, y/t(u,u,v), z/v\}$ is not.

\subsection{Optimal abstract operators}

All the above abstract domains are equipped with the optimal abstract operators for unification, matching and the other required operations. \cite{AmatoS09-tplp, AmatoS24-tplp} implement backward unification through matching, exploiting the property that the exit substitution is always more instantiated than the call substitution. 
Analyses based on matching have been shown to be strictly more precise than analyses that do
not use matching.
Moreover, \cite{AmatoS24-tplp} show that when using the optimal operators, the overall analysis is strictly more precise than the analysis performed with the standard operators. Surprisingly, the optimal operators are able to improve not only 
sharing and linearity information, but also groundness.

\section{Implementation}
\label{sec:implementation}
We have implemented the abstract domains \Sharing, \ShLin, \KingShLin, together with their operators, within the \PLAI analyzer.  All our code is publicly available on GitHub under the LGPL-3.0 license at \url{https://github.com/CLAI-UdA/ciaopp}. 

\subsection{Structure of the code}


\PLAI is written in Prolog and completely modular: each abstract domain is implemented in its own module and communicates with the analyzer engine through a well-defined interface. 
However, this interface is quite extensive and, to reduce the amount of code required for each abstract domain, we implemented a \emph{compatibility layer} in the module \module{as_template} which translates the standard interface into more elementary operations.

For example, consider the \prolog{call_to_entry} predicate, part of the abstract domain interface, corresponding to what we have called forward unification. It is defined as follows:
\begin{prologcode}
:- pred call_to_entry(+Sv, +Sg, +Hv, +Head, +ClauseKey, +Fv, +Proj,
                      -Entry, -ExtraInfo)
\end{prologcode}
where
\begin{itemize}
    \item \prolog{Sg} is the goal we are going to analyze;
    \item \prolog{Sv} is the set of variables occurring in \prolog{Sg};
    \item \prolog{ClauseKey} is an identifier for the clause we are using to analyze \prolog{Sg};
    \item \prolog{Head} is the head of the clause;
    \item \prolog{Hv} is the set of variables occurring in \prolog{Head};
    \item \prolog{Fv} are the variables in the clause that do not occur in  \prolog{Head};
    \item \prolog{Proj} is the call substitution projected onto the variables in \prolog{Sv};
    \item \prolog{Entry} is the resulting entry substitution;
    \item \prolog{ExtraInfo} may contain additional information that is passed later to the backward unification phase.
\end{itemize}
Our implementation of this predicate is the following:
\begin{prologcode}
call_to_entry(_Sv, Sg, Hv, Head, _ClauseKey, Fv, Proj, Entry, (Unif, Entry0)) :-
   unifiable_with_occurs_check(Sg, Head, Unifier),
   augment(Proj, Hv, ASub),
   mgu(ASub, Hv, Unifier, Entry0),
   project(Entry0, Hv, Entry1),
   augment(Entry1, Fv, Entry).
\end{prologcode}
It combines the predicates \prolog{augment}, \prolog{mgu} and \prolog{project}, implemented in each of our domains, with \prolog{unifiable_with_occurs_check} which is a built-in predicate in \Ciao.


The compatibility layer also avoids duplicating code that is syntactically identical across different domains. For example, the special value \texttt{\$bottom}, which denotes an empty abstract substitution (corresponding to $\bot$ in Section~\ref{sec:sharing}), is handled largely within this layer. The same applies to most of the code for analyzing Prolog built-ins.
Apart from the compatibility layer, there are two other modules that contain code shared by all our domains: \module{as_aux} contains generic utility code, while \module{as_bags} contains an implementation of finite multisets. 

Finally, the three domains are implemented in the modules \module{as_sharing}, \module{as_shlin} and \module{as_shlin2}. Although \CiaoPP already provides several implementations of \Sharing, we have devised a new one to serve as a baseline for the other domains in our collection. 
The behavior of the analysis (such as choosing between matching or unification for backward unification, or between optimal and standard operators) can be configured through specific \CiaoPP flags.

The built-in \PLAI domains for sharing analysis are implemented natively in \PROLOG, using sets (ordered lists) and a shallow encoding of logic variables. We adopt the same approach in our implementations. All code is fully annotated with \emph{type} and \emph{mode} information using the annotation mechanism provided by the \Ciao system. For example, this is the complete specification of the predicate computing the lubs of abstract substitutions:
\begin{prologcode}
:- pred compute_lub(+ListASub, -LubASub)
   : list_nonempty(asub) * ivar => asub(LubASub)
   + (not_fails, is_det).
\end{prologcode}
The specification states that \prolog{compute_lub} is always called with a non-empty list of abstract substitutions as its first argument, and an independent variable as a second argument, 
which is bound to an abstract substitution on exit. The predicate is also declared to be deterministic and not to fail (when called under these preconditions). These annotations may be checked at run-time, and have proved very useful during development, both for catching bugs in our code and for ensuring a correct understanding of how the \CiaoPP interface is expected to work. In addition, most modules include a comprehensive test suite. 


\subsection{From the algorithm to the implementation}

Moving from the algorithmic descriptions given in our papers to their actual implementation has not always been straightforward, and has required careful consideration.

\begin{figure}
\scalebox{0.90}{
\begin{tikzpicture}
    \node at (4cm,0cm) {Selected goal: $q(t(Y), Y)$ \qquad Clause: $q(t(A), B) \leftarrow \cdots$ \qquad $\mathcal E = \{q(t(Y),Y) = q(t(A),B)\}$};
    \node (cs) at (0,-1cm) [substitution] {Call substitution};
    \node [below of=cs] {\small
        \parbox{5.2cm}{
            \centering
            $p(X, Y), \colorbox{yellow}{q(t(Y), Y)}, r(Z)$\\[1ex]
            $\theta^{in}=\{X/f(\_0), Y/f(\_0), Z/\_1\}$
        }
    };
    \node (es) at (8cm, -1cm) [substitution] {Entry substitution};
    \node [below of=es] {\small
        \parbox{3.9cm}{
            \centering
            $\colorbox{yellow}{q(t(A), B)} \leftarrow \cdots$\\[1ex]            
            $\theta= \{ A/f(\_0), B/f(\_0) \}$
        }
    };
    \draw [->] (cs) edge
        node[above] {$\mgu\big(\theta^{in}, \mathcal E\big)$}
        node[below] {\color{red}{\prolog{call_to_entry}}}
    (es);
    \node (ees) at (0, -4cm) [substitution] {Answer substitution};
    \node [below of=ees] {\small
        \parbox{7.2cm}{
            \centering
            $p(X, Y), \colorbox{yellow}{q(t(Y), Y)}, r(Z)$\\[1ex]
            $\theta^{out}=\{X/f(g(\_2,\_2)), Y/f(g(\_2,\_2)), Z/\_1\}$
        }
    };
    \node (as) at (8cm, -4cm) [substitution] {Exit substitution};
    \node [below of=as] {\small
        \parbox{5.8cm}{
            \centering
            $q(t(A), B) \leftarrow \colorbox{yellow}{$\cdots$}$\\[1ex]
            $\theta'=\{A/f(g(\_2, \_2)), B/f(g(\_2, \_2))\}$
        }
    };
    \node (prime) at (4, -6cm) [substitution] {Prime};
    \node [below of=prime] {\small
        \parbox{6.6cm}{
            \centering
            $\colorbox{yellow}{q(t(Y), Y)}$\\[1ex]
            $\theta''=\{Y/f(g(\_2,\_2))\}$
        }
    };
    \draw [decorate,
        decoration={
            text along path,      
            text color=red,
            text={|\tt|\detokenize{exit_to_prime}},
            text align=center,
            raise=-2ex
        }
    ] (prime.north) to [bend left=40](as.west);
    \draw [<-] (prime.north) to [bend left=40] node[above, pos=0.7, sloped]{$\match(\theta', \mgu(\theta^{in},\mathcal E))$} (as.west);
    \draw [decorate,decoration={
        text along path,      
        text color=red,
        text={|\tt|extend},
        text align=center,
        raise=-2ex
    }] (ees.east)  to [bend left=40] (prime.north);
    \draw [<-] (ees.east)  to [bend left=40] node[above, pos=0.2, sloped]{$\match(\theta'', \theta^{in})$} (prime.north);    
    \draw [->] (es) -- ($ (es.east)+(+0.5cm,+0cm) $) --  node[above,rotate=-90]{body analysis} ($ (as.east)+(+0.5cm,+0cm) $) --  (as);
\end{tikzpicture}
}
\caption{\label{fig:plai-flow}Flow of the analysis in \PLAI.}
\end{figure}
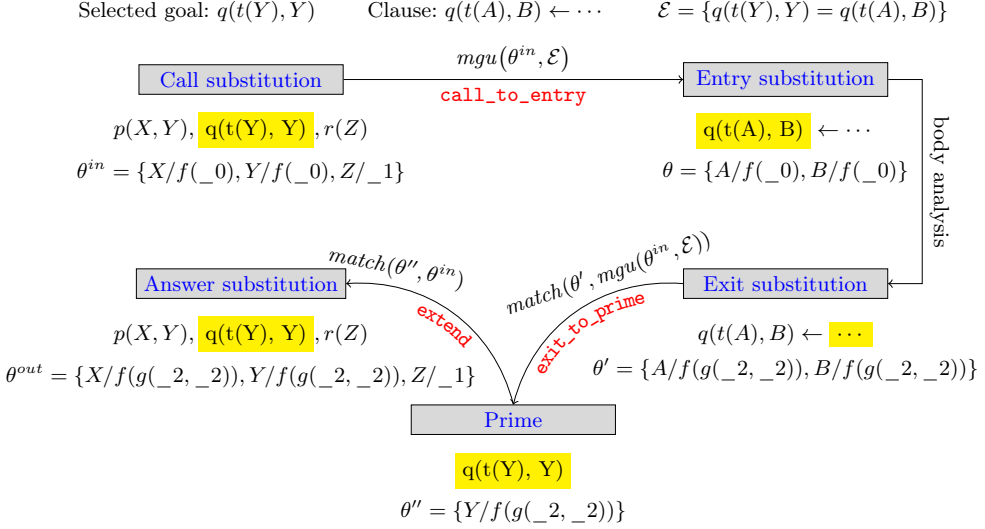

First of all, while \PLAI has the predicate \prolog{call_to_entry} which corresponds directly to forward unification, it splits backward unification into two distinct steps. First, the exit substitution is  projected onto the variables of the chosen atom by the predicate \prolog{exit_to_prime}, it is then extended to all the variables of the original goal by the predicate \prolog{extend}, incorporating the information on the other variables coming from the call substitution.
This process is depicted in Figure~\ref{fig:plai-flow}, which we briefly describe here. The procedure $\prolog{call_to_entry}$ computes the $\mgu$ between $\theta^{in}$ and $\mathcal E$, the equation connecting the selected atom $q(t(Y), Y)$ and the clause head $q(t(A), B)$, yielding $\bar \theta = \mgu(\theta^{in}, \mathcal E) = \{X/f(\_0),Y/f(\_0),Z/\_1,A/f(\_0),B/f(\_0)\}$. This is then projected onto the variables of the clause to obtain the entry substitution $\theta$. After analyzing the body, $\theta$ is further instantiated, yielding the exit substitution $\theta'$. 

Now \prolog{exit_to_prime} computes the $\mgu$ between $\theta'$ and $\bar\theta$, and projects the result onto the variables of the selected goal, obtaining $\theta''$, called the \emph{prime}. Note that in this step we are sure that variables in $\theta'$ are not further instantiated, hence unification may be replaced by matching. Using matching instead of unification (which is irrelevant in the concrete) improves precision in the abstract. 
Finally, \prolog{extend} propagates the effect of the prime to the entire initial goal, by unifying it with the call substitution. Again, since variables in the prime are not going to be further instantiated, unification may be replaced by matching.

Two observations are worth noting. First, the \texttt{extend} step is conceptually redundant: the final projection in \prolog{exit_to_prime} could be performed directly over all variables of the initial goal, thus producing the answer substitution in a single step. However, this would require modifying the \PLAI engine, which we do not intend to do.
Second, our $\mgu$ and $\match$ operations are designed so that $\mgu$ takes an abstract substitution and a concrete set of equations, whereas $\match$ takes two abstract substitutions. 
This makes the implementation of \prolog{extend} using only $\mgu$ a bit cumbersome, since it should operate on two abstract substitutions.

More importantly, the definitions of $\mgu$ and $\match$ given in our papers are expressed in a declarative style. It is relatively easy to implement these operators following the definitions using a generate-and-test approach, and we have adopted this strategy in several cases. However, there are some cases in which this approach is too slow, making it necessary to devise more carefully optimized implementations. 

As an illustrative example, consider the definition of the matching operator for \ShLin shown in Figure~\ref{def:match}. Rather than discussing the details of the definition (for which we refer to \cite{AmatoS10-tplp}), we use it here to highlight some of the issues that arise in its implementation.
The crucial step in the definition is the computation of $S''_0$. This involves selecting a sharing group $B \in S_1$ and a subset of sharing groups $X \subseteq S''_2$, and then checking whether the conditions $B \cap U_2 = \left(\bigcup X\right) \cap U_1$ and $L_1 \cap nl(X)=\emptyset$ are satisfied.

A straightforward implementation would follow a generate-and-test strategy that mirrors the formal definition. However, this approach requires enumerating a large number of candidate subsets $X$, whose number is exponential in the size of $S''_2$, which is itself potentially exponential in the number of program variables. To obtain a practically usable implementation, it is therefore necessary to reduce the search space.

First, once we fix $B \in S_1$, a necessary condition for satisfying $B \cap U_2 = \left(\bigcup X\right) \cap U_1$ is that every $C \in X$ satisfies $C \cap U_1 \subseteq B \cap U_2$. Hence, we can restrict the search to subsets $X \subseteq \{ C \in S''_2 \mid C \cap U_1 \subseteq B \cap U_2 \}$. Additionally, instead of generating $X$, checking constraints and then producing the pair $\left\langle B \cup \bigcup X, L_2  \setminus nl(X) \setminus \bigcup (X \cap \bar S) \right\rangle$, we have devised a procedure that incrementally builds the resulting pairs, without explicitly generating the candidates $X$ at all. In this way, although the theoretical complexity of the procedure does not decrease, the implementation is generally much faster.

\begin{figure}
{\footnotesize
\fbox{\begin{minipage}{0.98\textwidth}
\[
  \matchshl([S_1,L_1, U_1],[S_2,L_2,U_2])= \left[s(S'_0 \cup S''_0), l(S'_0 \cup S''_0), U \right]
\]
\begin{align*}
  U & = U_1 \cup U_2 & nl(X) & = \{ v \in \varset \mid \exists B_1, B_2 \in X, B_1 \neq B_2, v \in B_1 \cap B_2 \}\\
  S'_2 & =\{ B \in S_2 \mid B \cap U_1 = \emptyset \} & s(H)  & = \left\{ B \mid \langle B, L \rangle \in H \right\}\\
  S''_2 & =S_2 \setminus S'_2 & l(H)  & = \bigcap \left\{ L_1 \cup L \cup (U \setminus B) \mid \langle B,L \rangle \in H \right\} \\
  \bar S & = \{B \in S''_2 \mid B \cap L_1 = \emptyset \} & S'_0  & = \bigl\{ \langle B, L_2 \rangle \mid B \in S'_2 \bigr\}\\
  & & S''_0 & = \Big\{ \left\langle B \cup \bigcup X, L_2  \setminus nl(X) \setminus \bigcup (X \cap \bar S) \right\rangle \Bigm| B \in S_1,\\
  &&      & \qquad X \subseteq S''_2,  B \cap U_2 = \left(\bigcup X\right) \cap U_1, L_1 \cap nl(X)=\emptyset \Big\} 
\end{align*}
\end{minipage}}
}
\caption{\label{def:match}The definition of the match operator in \ShLin, for $[S_1,L_1,U_1]$ and $[S_2,L_2,U_2] \in \ShLin$.}
\end{figure}

\section{Experimental evaluation}
\label{sec:experimental-evaluation}

In order to benchmark both the precision and the efficiency of our domains,  we used the \SWI benchmark suite\footnote{\url{https://github.com/SWI-Prolog/bench}, commit \href{https://github.com/SWI-Prolog/bench/commit/f3269642bc6b86fe48805db6e7634c3d05bd2618}{f326964}}. It consists of 35 Prolog programs, collected by the \SWI organization but running on many Prolog systems. Two programs were removed from the suite: \module{det.pl}, which does not run in \Ciao, and \module{queens_clpfd.pl}, which relies on CLP(FD) features that are not supported in our setting. As a result, we analyzed a total of 33 programs. They are listed in Table~\ref{tab:programs-size} together with some basic size metrics.

\begin{table}
\centering
\footnotesize
\begin{filecontents*}{program-sizes.csv}
program,size,predicates,clauses,totvar,maxvar
boyer,11232,25,135,312,7
browse,2831,16,32,119,12
chat_parser,25824,158,516,1650,23
crypt,1834,9,27,64,19
derive,975,5,14,40,5
divide10,857,3,12,38,5
eval,639,5,6,11,4
fast_mu,3359,9,18,102,13
fib,1109,2,4,7,6
flatten,5773,27,58,221,13
log10,882,3,12,38,5
meta_qsort,2829,8,26,62,7
moded_path,2429,6,21,26,6
mu,945,9,17,34,6
nand,19927,42,138,714,44
nreverse,616,4,6,10,4
ops8,835,3,12,38,5
perfect,2061,9,14,40,5
pingpong,480,4,7,5,2
poly_10,3233,12,33,106,8
prover,2352,10,33,92,9
qsort,772,4,7,20,7
queens_8,1753,7,12,30,5
query,1754,6,55,11,6
reducer,11743,43,122,459,17
sendmore,1178,4,22,17,11
serialise,1108,8,14,45,7
sieve,1173,6,9,21,5
simple_analyzer,14154,67,143,662,48
tak,608,3,4,15,10
times10,853,3,12,38,5
unify,6945,15,63,291,14
zebra,1608,7,12,128,78
\end{filecontents*}
\csvreader[
  respect underscore,
  tabular=l*{5}r,
  table head={
    \toprule
    program & size & \#predicates & \#clauses & total vars & max vars\\
    \midrule
  },
  table foot=\bottomrule,
]{program-sizes.csv}{
  program=\prog,
  size=\sizep,
  predicates=\pred,
  clauses=\clauses,
  totvar=\totvar,
  maxvar=\maxvar
}{%
    \prog & \sizep & \pred & \clauses & \totvar & \maxvar
}
\caption{\label{tab:programs-size}\small Benchmark programs and corresponding size metrics. The column \emph{max vars} is the maximum number of variables in a clause.}
\end{table}

Each program was analyzed in 15 different configurations.
These include 
the built-in \Sharing domain provided by \PLAI (\module{share_amgu}), the built-in $\Sharing \times \Free \times \Lin$ (\module{shfrlin_amgu}), 
as well as our implementations of \Sharing, \ShLin and \KingShLin with different options. 
In the following, we report only a representative subset of these configurations.
The configuration referred to as \emph{base} uses the standard (as described in the literature) $\mgu$ for both forward and backward unification. In particular, the standard operators used for \ShLin are the ones by \cite{HansW92-tr}, including the independence checks that \cite{HoweK03-tplp} and \cite{HillZB04-tplp} showed to be removable. The configuration \emph{match} employs the standard $\mgu$ for forward unification but the (optimal) matching operator for backward unification. Finally, the configuration \emph{optimal} uses optimal operators for both $\mgu$ and $\match$. 

The built-in domains \module{share_amgu} and \module{shfrlin_amgu} rely on specialized implementations that likely do not correspond to any of the configurations considered in this work. To the best of our knowledge, these implementations have not been fully described in the literature, at least in their current form.
Note that \module{shfrlin_amgu} also tracks freeness information. This additional component may both improve precision 
and introduce extra computational overhead, which should be taken into account when comparing performance.

All benchmarks were executed on a machine equipped with an Intel i5-13600K CPU with 32 GB of RAM, running Ubuntu Linux 25.10 and \Ciao \texttt{1.25.0-m1}. We measured both execution time and precision, according to the following procedure:
\begin{itemize}
\item \CiaoPP was instructed to analyze the main predicate of each program;
\item other predicates were analyzed only as a consequence of this initial call;
\item a timeout of 2 minutes was imposed for each analysis;
\item the virtual address space of the analyzer was limited to 24GB of RAM.
\end{itemize}
The 24GB memory limit was introduced to reliably detect out-of-memory conditions. In preliminary experiments, we relied on the operating system’s OOM killer; however, this approach proved inconsistent, as it occasionally terminated not only the analyzer but also the script orchestrating the benchmark execution.   

Following this procedure, we were able to fully analyze all benchmarks, with the exception of six critical programs, which failed under at least one configuration. These programs are summarized in Figure~\ref{fig:critical}. 
There are two cases, namely  \module{reducer} with \ShLin and \module{flatten} with \KingShLin, where increasing precision of abstract operators leads to non-termination, and two cases, namely \module{chat_parser} with \ShLin, \module{reducer} with \KingShLin and \module{unify} with both \Sharing and \ShLin, where increasing precision allows 
the analysis to complete within the timeout. 

As expected, the program metric most strongly correlated with analysis failures is the maximum number of variables in a clause (\texttt{max vars}), as it directly affects the size of abstract objects. In fact, the six programs in Figure~\ref{fig:critical} are all among the top nine programs with the highest values of the \texttt{max vars} metric.
This observation is confirmed by computing the Pearson correlation coefficient between the indicator function of the timeout/out-of-memory condition and each metric. The highest correlation is consistently obtained with \texttt{max vars} across all abstract domains. For instance, for \ShLin with optimal operators, the correlation is 0.72 with \texttt{max vars}, while it remains below 0.30 for all other metrics.

\begin{figure}
    \begin{tabular}{l|c|c|c|c|c|c|c|c|c|c|c|}
        Program & \multicolumn{3}{c|}{\KingShLin} & \multicolumn{3}{c|}{\ShLin} & \multicolumn{3}{c|}{\Sharing} & \multicolumn{2}{c|}{built-in}\\[1ex]
        & \rotatebox{90}{optimal} & \rotatebox{90}{match} & \rotatebox{90}{base}  & \rotatebox{90}{optimal} & \rotatebox{90}{match} & \rotatebox{90}{base}  & \rotatebox{90}{optimal} & \rotatebox{90}{match} & \rotatebox{90}{base} & \rotatebox{90}{\Sharing} &  \rotatebox{90}{\ShFrLin} \\
        \hline
        \texttt{chat\_parser} & \xmark & \xmark & \xmark &  \checkmark & \xmark & \xmark & \xmark  & \xmark & \xmark & \xmark & \xmark \\
        \texttt{flatten}& \xmark & \xmark & \checkmark &  \checkmark & \checkmark & \checkmark & \checkmark  & \checkmark & \checkmark & \checkmark & \checkmark \\
        \texttt{reducer} & \checkmark & \checkmark & \xmark &  \xmark & \checkmark & \xmark & \checkmark  & \checkmark & \xmark & \xmark & \checkmark \\
        \texttt{simple\_analyzer} & \xmark & \xmark & \xmark &  \xmark & \xmark & \xmark & \xmark  & \xmark & \xmark & \xmark & \xmark \\
        \texttt{unify} & \checkmark & \checkmark & \checkmark &  \checkmark & \checkmark & \xmark & \checkmark  & \checkmark & \xmark & \checkmark & \checkmark \\
        \texttt{zebra} & \xmark & \xmark & \xmark &  \xmark & \xmark & \xmark & \xmark  & \xmark & \xmark & \xmark & \xmark \\
    \end{tabular}
    \caption{\label{fig:critical}Critical programs, causing either a timeout or an out-of-memory condition (both \xmark).}
\end{figure}

\newlength{\plotsize}
\setlength{\plotsize}{0.38\textwidth}

\begin{figure}
    \includegraphics[width=\plotsize]{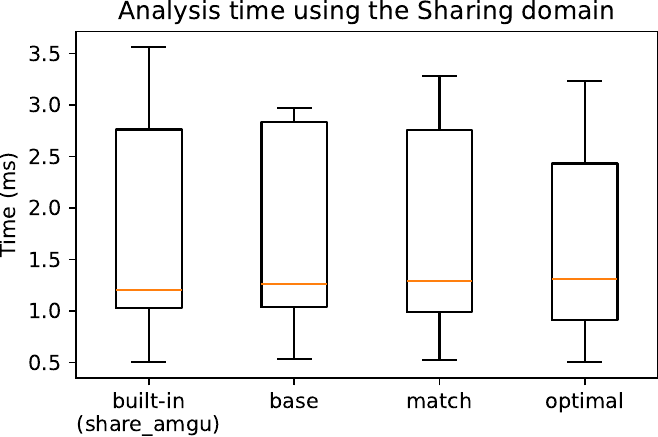}
    \hspace{1cm}
    \includegraphics[width=\plotsize]{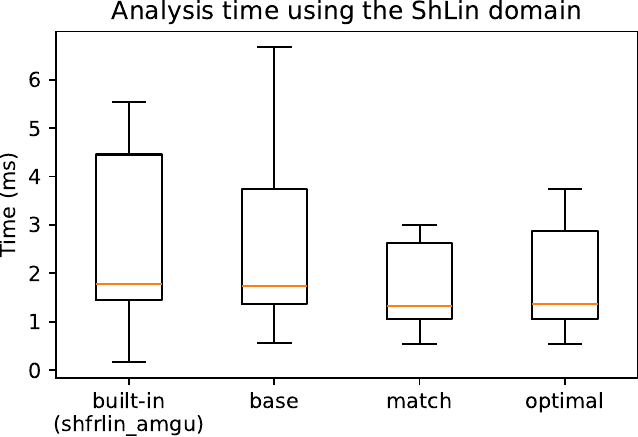}

    \vspace{3ex}
    \includegraphics[width=\plotsize]{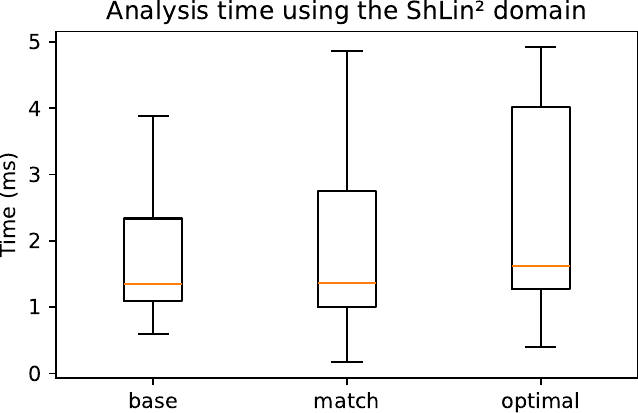}
    \hspace{1cm}
    \includegraphics[width=\plotsize]{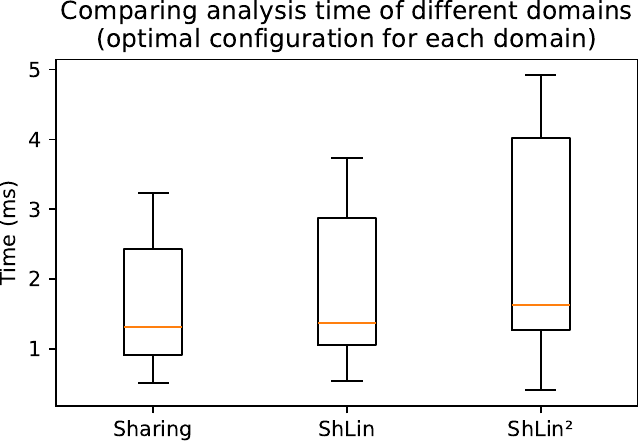}
    \caption{\label{fig:time}Analysis time for non-critical programs.}
\end{figure}

Figure~\ref{fig:time} reports the execution time for all non-critical programs, while Table~\ref{tab:report-time} provides the complete set of results.
For each domain we show a boxplot comparing different configurations, followed by a final boxplot comparing the three domains under optimal operators. As expected, more complex domains lead to higher execution times. However, the increase is negligible when comparing \ShLin with \Sharing, while it is much more visible for \KingShLin. This behavior is expected, since the size of abstract objects for the latter is much larger. Analyzing the boxplots for each domain confirms what we already observed for the critical programs: while more complex operators may increase analysis time, this effect does not occur systematically, the increase is generally negligible and sometimes the opposite happens.


If we exclude timeout and out-of-memory cases, the correlation between analysis time and the considered metrics becomes less clear. In particular, \texttt{max vars} generally shows a low correlation, while \texttt{predicates} tends to exhibit higher values. However, these trends vary significantly across the different abstract domains.

\begin{figure}
    \includegraphics[width=\plotsize]{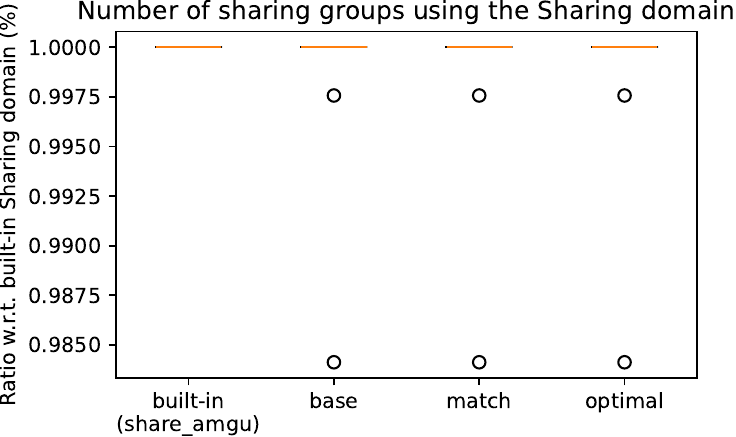}
    \hspace{1cm}
    \includegraphics[width=\plotsize]{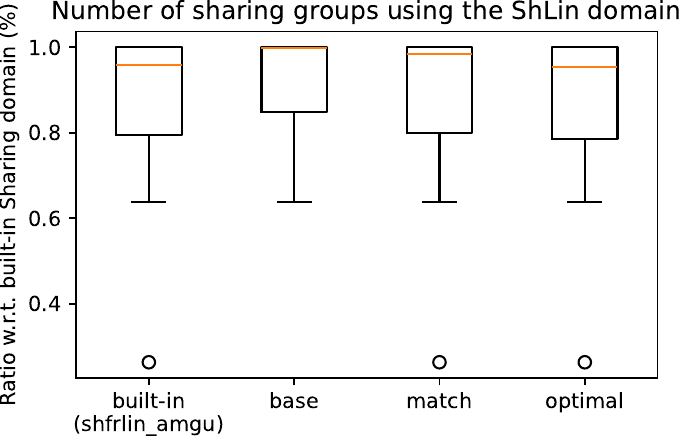}

    \vspace{3ex}

    \includegraphics[width=\plotsize]{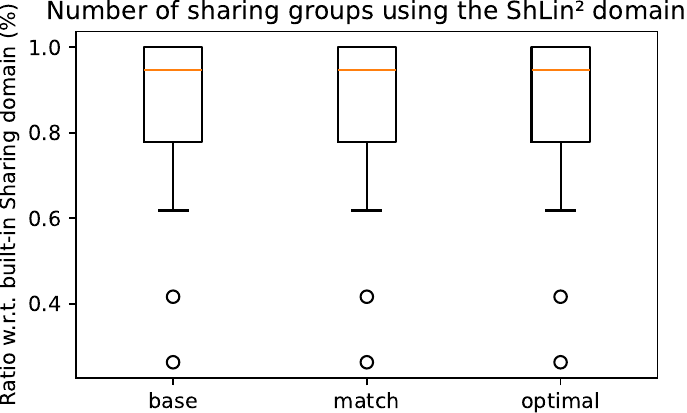}
    \hspace{1cm}
    \includegraphics[width=\plotsize]{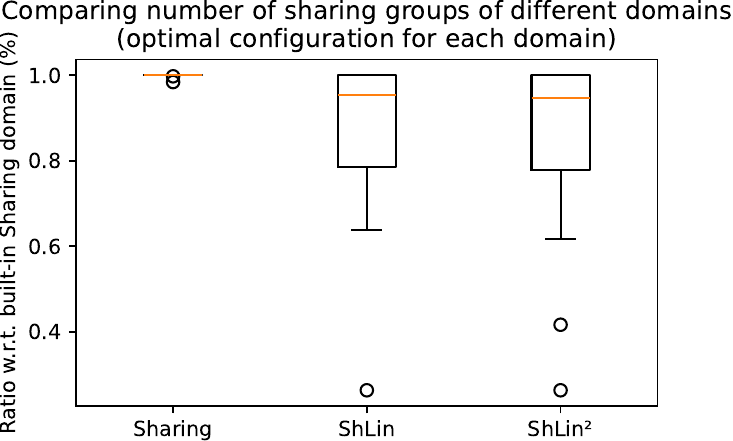}
    \caption{\label{fig:precision}Precision with respect to the sharing property.}
\end{figure}

Figure~\ref{fig:precision} reports the precision of the three domains with respect to the sharing property. For each program and configuration, we counted the number of sharing groups in the annotations generated by \CiaoPP. For program points where \CiaoPP reports multiple analysis results, corresponding to different preconditions, we only counted the sharing groups appearing in the alternative with the smallest number of sharing groups. Although this is a coarse measure of precision, it provides a useful indication of overall trends. 
It shows that the use of matching and optimal operators leads to a clear increase in precision for the \ShLin domain, while the improvement is more limited for \Sharing and \KingShLin. Moreover, more expressive domains yield more precise results.

\begin{figure}
    \includegraphics[width=\plotsize]{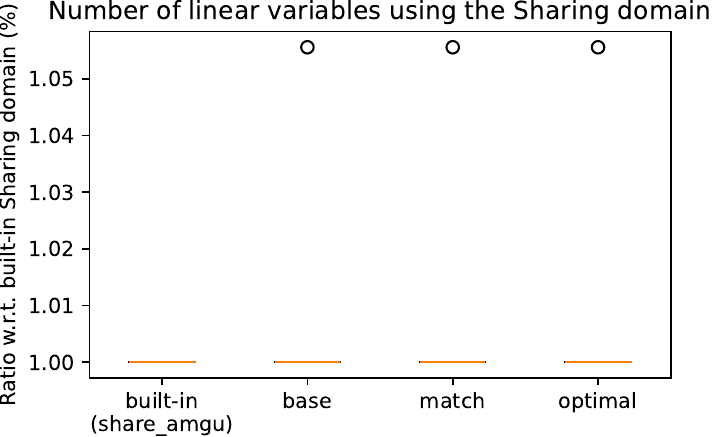}
    \hspace{1cm}
    \includegraphics[width=\plotsize]{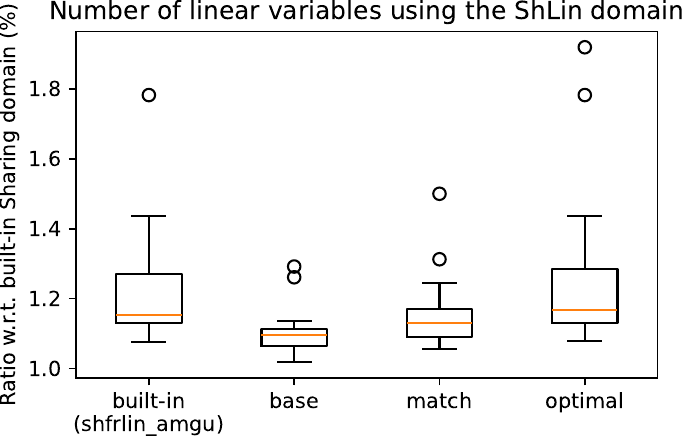}

    \vspace{3ex}
    \includegraphics[width=\plotsize]{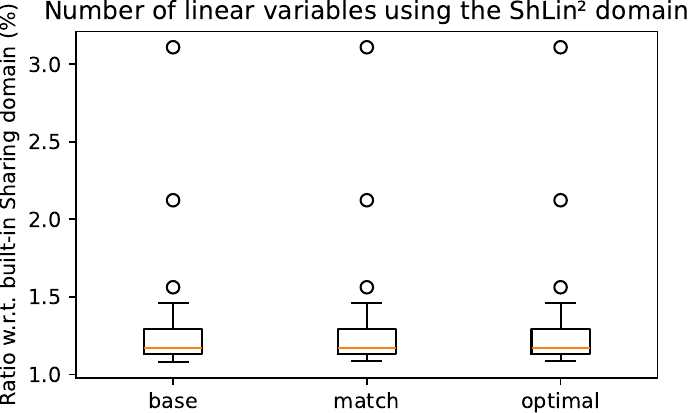}
    \hspace{1cm}
    \includegraphics[width=\plotsize]{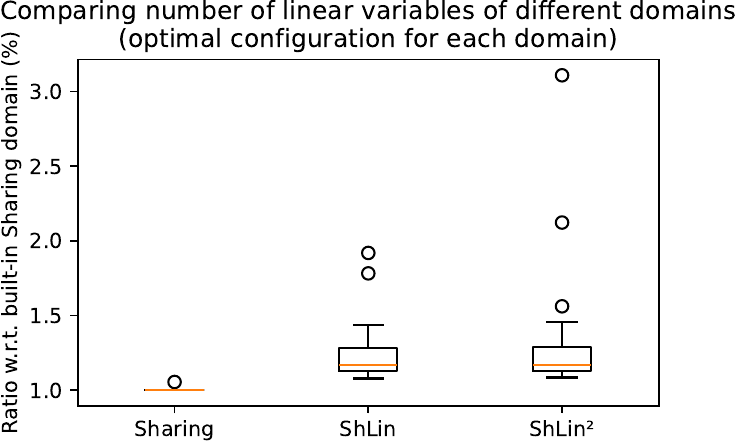}
    \caption{\label{fig:linear-precision}Precision with respect to the linearity property.}
\end{figure}

Figure~\ref{fig:linear-precision} reports the precision with respect to the linearity property, measured by counting the number of linear variables, and computing the maximum in case of multiple annotations. The results are similar to those obtained for the sharing property.
We do not report boxplots for groundness, since almost all configurations yield identical results.

The complete set of detailed results is provided in Table~\ref{table:precision}, which shows that \KingShLin with matching and optimal operators yields the most precise results among all the considered domains, with the only exception of the programs \module{chat_parser} and \module{flatten}, for which the analysis does not terminate. 
%
Since we measure sharing precision as the number of sharing groups, this also provides an indication of the size of abstract objects. As expected, in several cases, optimal operators lead to smaller abstract objects. This effect concerns the sharing component only, which is the dominant one, as the number of sharing groups may grow exponentially with the number of variables.

%
Finally, our implementation of the \Sharing domain, when used with the most precise configuration, consistently outperforms the built-in \Sharing implementation in \CiaoPP; in some cases, it succeeds in terminating analyses for which the built-in implementation does not terminate.




\begin{sidewaystable}[hp]
\centering
\tiny
\setlength{\tabcolsep}{3pt}
\begin{filecontents*}{report_time_3.csv}
program,as_shlin2_opt,as_shlin2_opt_mgu,as_shlin2_noopt,as_shlin2_noopt_mgu,as_shlin_opt_opt,as_shlin_opt,as_shlin_noindcheck,as_shlin_noopt,as_shlin_opt_mgu,as_shlin_noindcheck_mgu,as_shlin_noopt_mgu,as_sharing_opt,as_sharing_noopt,as_sharing_opt_mgu,as_sharing_noopt_mgu,share,shfrlin
boyer,37.121,57.096,23.932,20.792,51.96,41.358,33.216,33.122,1242.125,253.599,254.62,28.371,31.557,242.264,248.382,25.141,28.793
browse,280.081,250.71,272.285,10.486,29.836,32.216,31.548,31.894,3369.263,257.422,256.255,29.768,586.998,249.751,259.373,15.459,16.814
chat_parser,OOM,TO,OOM,TO,62322.589,24305.137,TO,TO,TO,TO,TO,TO,TO,TO,TO,TO,TO
crypt,3.972,5.354,4.863,3.094,2.849,3.043,2.891,2.989,2.802,3.209,3.311,2.438,3.019,1.819,2.821,2.541,4.522
derive,1.889,2.39,1.368,1.461,1.439,1.629,1.491,1.419,2.316,1.734,2.122,1.341,1.352,1.467,1.429,1.415,1.902
divide10,1.474,1.94,1.027,1.133,1.079,1.139,1.055,1.069,1.859,1.763,1.736,0.989,1.054,1.132,1.093,1.049,1.471
eval,0.776,0.853,0.633,0.689,0.689,0.707,0.676,0.709,0.81,0.694,0.688,0.637,0.673,0.677,0.697,0.674,0.785
fast_mu,4.353,5.777,1.626,2.135,2.697,2.969,2.859,1.909,3.549,3.254,3.316,2.341,2.408,2.602,2.312,2.8,4.58
fib,0.676,0.75,0.573,0.625,0.598,0.573,0.551,0.579,0.858,0.837,0.824,0.519,0.527,0.021,0.56,0.567,0.163
flatten,OOM,305.321,OOM,139.053,55.569,44.193,122.589,124.271,6029.845,1158.178,1164.468,104.683,112.588,990.709,1008.485,148.937,46.461
log10,1.342,1.772,0.986,1.087,1.038,1.069,1.064,1.049,1.845,1.662,1.754,0.976,1.002,1.055,1.084,1.016,1.423
meta_qsort,42.202,44.842,27.717,16.639,8.479,7.645,11.238,11.137,28.358,14.497,14.622,6.808,6.15,8.607,9.199,7.752,9.402
moded_path,1.626,1.957,1.216,1.31,1.264,1.294,1.26,1.217,1.387,1.29,1.326,1.088,1.143,1.184,1.181,1.175,1.777
mu,2.456,3.208,1.65,1.757,1.84,1.9,1.741,1.333,2.055,1.931,1.935,1.634,1.623,1.781,1.789,1.794,2.536
nand,58.464,68.381,22.331,25.946,31.262,32.417,93.743,93.841,88.356,76.39,82.523,85.971,86.652,48.461,48.871,43.73,39.258
nreverse,0.935,1.018,0.685,0.777,0.723,0.732,0.706,0.725,0.794,0.76,0.721,0.666,0.652,0.758,0.747,0.685,0.918
ops8,1.414,1.925,1.067,1.095,1.091,1.156,1.076,1.055,1.849,1.7,1.717,0.984,0.988,1.111,1.097,1.057,1.472
perfect,1.797,2.269,1.467,1.579,1.53,1.524,1.479,1.143,1.664,1.628,1.62,1.355,1.406,1.496,1.464,1.07,1.925
pingpong,0.594,0.627,0.168,0.59,0.537,0.551,0.53,0.545,0.549,0.536,0.559,0.503,0.522,0.521,0.532,0.509,0.554
poly_10,4.918,6.71,3.317,3.884,3.734,3.846,5.293,5.231,5.299,5.587,6.67,3.234,3.281,3.676,2.847,3.561,5.533
prover,4.072,5.606,2.543,2.079,2.908,3.012,3.135,2.272,4.995,4.085,4.152,2.427,2.504,2.977,2.972,2.725,4.387
qsort,1.201,1.671,0.939,1.028,0.986,1.01,1.405,1.435,1.147,1.544,1.603,0.855,0.895,0.995,0.997,0.95,1.358
queens_8,1.502,1.587,1.164,1.251,1.246,1.284,1.271,1.281,1.424,1.406,1.426,1.121,1.157,1.218,1.23,1.205,1.68
query,1.968,2.075,1.372,1.342,1.362,1.395,1.314,1.331,1.735,1.412,1.411,1.329,1.296,1.248,0.988,1.28,1.784
reducer,32.608,TO,16.821,TO,TO,29509.867,70973.742,71091.251,TO,TO,TO,70327.287,71882.298,TO,TO,TO,44293.176
sendmore,0.885,2.041,1.411,1.655,1.482,1.504,1.465,1.436,1.812,1.748,1.766,1.311,1.288,1.437,1.406,1.277,1.813
serialise,3.093,4.728,2.954,2.537,3.664,4.28,16.78,16.348,134.471,43.438,43.098,12.953,13.844,37.129,40.664,13.206,3.471
sieve,1.45,1.694,1.218,1.278,1.274,1.295,1.231,1.253,1.381,1.329,1.325,1.179,1.185,1.241,1.263,1.205,1.55
simple_analyzer,TO,TO,OOM,TO,TO,TO,TO,TO,TO,TO,TO,TO,TO,TO,TO,TO,TO
tak,0.404,1.163,0.68,0.754,0.685,0.307,1.002,0.9,1.215,1.143,1.147,0.614,0.621,0.706,0.777,0.618,0.884
times10,1.39,1.856,1.019,1.114,1.108,1.215,1.068,1.048,1.893,1.724,1.724,0.677,1.001,1.1,1.118,1.074,1.478
unify,7536.688,588.872,7474.641,207.981,106.249,129.316,973.752,974.414,TO,TO,TO,949.814,977.158,TO,TO,1862.961,157.956
zebra,TO,TO,OOM,TO,OOM,OOM,TO,TO,TO,TO,TO,TO,TO,TO,TO,TO,TO
\end{filecontents*}
\csvreader[
  respect all,
  tabular=l*{15}{r},
  table head={
    \toprule
    program &
    shlin2+ & shlin2+M & shlin2 & shlin2M &
    shlin++ & shlin+ &  shlin &
    shlin+M & shlinM &
    sharing+ & sharing & sharing+M & sharingM &
    share & shfrlin \\
    \midrule
  },
  late after line=\\,
  table foot=\bottomrule
]{report_time_3.csv}{
  program=\prog,
  as_shlin2_opt=\cA,
  as_shlin2_opt_mgu=\cB,
  as_shlin2_noopt=\cC,
  as_shlin2_noopt_mgu=\cD,
  as_shlin_opt_opt=\cE,
  as_shlin_opt=\cF,
  as_shlin_noopt=\cH,
  as_shlin_opt_mgu=\cI,
  as_shlin_noopt_mgu=\cK,
  as_sharing_opt=\cL,
  as_sharing_noopt=\cM,
  as_sharing_opt_mgu=\cN,
  as_sharing_noopt_mgu=\cO,
  share=\cP,
  shfrlin=\cQ
}{%
  \prog &
  {\cA} & {\cB} & {\cC} & {\cD} &
  {\cE} & {\cF} & {\cH} &
  {\cI} & {\cK} &
  {\cL} & {\cM} & {\cN} & {\cO} &
  {\cP} & {\cQ}
}
\caption{\label{tab:report-time}
Analysis time for all programs with the different abstract domains and configurations. \emph{OOM} denotes an out-of-memory condition, while \emph{TO} denotes a timeout.}
\end{sidewaystable}

\begin{table}
\centering
\tiny
\setlength{\tabcolsep}{2pt}
\begin{adjustbox}{max width=\textwidth}
\begin{filecontents*}{report_precision_best_5.csv}
property,program,as_shlin2(+),as_shlin2_opt_mgu,as_shlin2_noopt_mgu,as_shlin_opt_opt,as_shlin_opt,as_shlin_noopt,as_shlin_opt_mgu,as_shlin_noopt_mgu,as_sharing(+),as_sharing_mgu(+),share,shfrlin
mshare,boyer,296,296,296,321,321,437,437,437,479,479,479,437
mshare,browse,765,765,765,804,804,811,817,821,821,821,823,809
mshare,chat_parser,OOM,TO,TO,5804,5804,TO,TO,TO,TO,TO,TO,TO
mshare,crypt,224,224,224,224,224,235,235,235,235,235,235,235
mshare,derive,23,23,23,23,23,23,23,23,29,29,29,23
mshare,divide10,21,21,21,21,21,21,21,21,27,27,27,21
mshare,eval,8,8,8,8,8,8,8,8,8,8,8,8
mshare,fast_mu,97,97,97,97,97,116,116,116,116,116,116,97
mshare,fib,21,21,21,21,21,21,21,21,21,21,21,21
mshare,flatten,OOM,1335,1335,1343,1343,1511,1522,1524,1584,1584,1754,1495
mshare,log10,21,21,21,21,21,21,21,21,27,27,27,21
mshare,meta_qsort,62,62,62,62,62,62,62,62,62,62,63,62
mshare,moded_path,16,16,16,16,16,16,16,16,16,16,16,16
mshare,mu,13,13,13,13,13,13,13,13,16,16,16,13
mshare,nand,757,757,757,757,757,757,2879,2879,2879,2879,2879,757
mshare,nreverse,5,5,5,5,5,5,5,5,5,5,5,5
mshare,ops8,21,21,21,21,21,21,21,21,27,27,27,21
mshare,perfect,46,46,46,46,46,46,46,46,48,48,48,46
mshare,pingpong,9,9,9,9,9,9,9,9,9,9,9,9
mshare,poly_10,54,54,54,54,54,54,54,54,61,61,61,54
mshare,prover,36,36,36,36,36,36,36,36,38,38,38,36
mshare,qsort,14,14,14,14,14,14,14,14,14,14,14,14
mshare,queens_8,19,19,19,19,19,19,19,19,19,19,19,19
mshare,query,23,23,23,23,23,23,23,23,36,36,36,23
mshare,reducer,269,TO,TO,TO,15359,16736,TO,TO,16885,TO,TO,16919
mshare,sendmore,127,127,127,127,127,127,127,127,127,127,127,127
mshare,serialise,102,102,102,195,195,197,225,227,245,245,245,195
mshare,sieve,19,19,19,19,19,19,19,19,19,19,19,19
mshare,simple_analyzer,TO/OOM,TO,TO,TO,TO,TO,TO,TO,TO,TO,TO,TO
mshare,tak,38,38,38,38,38,38,38,38,38,38,38,38
mshare,times10,21,21,21,21,21,21,21,21,27,27,27,21
mshare,unify,1287,1367,1367,1365,1368,2487,TO,TO,2602,TO,2602,1375
mshare,zebra,TO/OOM,TO,OOM,OOM,OOM,TO,TO,TO,TO,TO,TO,TO
linear,boyer,383,383,383,335,335,152,173,150,118,118,118,173
linear,browse,376,376,376,299,299,259,223,211,177,177,177,289
linear,chat_parser,OOM,TO,TO,4261,4261,TO,TO,TO,TO,TO,TO,TO
linear,crypt,620,620,620,620,620,589,429,429,396,396,396,620
linear,derive,105,105,105,105,105,105,101,101,82,82,82,105
linear,divide10,101,101,101,101,101,101,97,97,80,80,80,101
linear,eval,42,42,42,42,42,42,41,41,34,34,34,42
linear,fast_mu,358,355,355,355,355,275,300,264,246,246,246,353
linear,fib,45,45,45,45,45,39,42,38,24,24,24,45
linear,flatten,OOM,378,378,379,377,247,279,226,143,143,141,400
linear,log10,101,101,101,101,101,101,97,97,80,80,80,101
linear,meta_qsort,134,134,134,105,105,80,93,80,76,76,72,105
linear,moded_path,55,55,55,52,52,44,47,43,39,39,39,51
linear,mu,78,78,78,78,78,78,77,77,65,65,65,78
linear,nand,3885,3885,3885,3885,3885,3621,3341,3329,3132,3132,3132,3885
linear,nreverse,22,22,22,22,22,19,20,19,17,17,17,22
linear,ops8,101,101,101,101,101,101,97,97,80,80,80,101
linear,perfect,143,143,143,143,143,140,133,131,107,107,107,143
linear,pingpong,20,20,20,20,20,14,16,14,11,11,11,20
linear,poly_10,288,288,288,288,288,274,276,267,234,234,234,288
linear,prover,260,257,257,257,257,247,246,240,222,222,222,255
linear,qsort,55,55,55,55,55,49,52,48,41,41,41,55
linear,queens_8,89,89,89,89,89,83,85,83,70,70,70,89
linear,query,62,62,62,62,62,58,50,48,39,39,39,62
linear,reducer,1216,TO,TO,TO,672,563,TO,TO,418,TO,TO,710
linear,sendmore,469,469,469,469,469,469,354,354,342,342,342,469
linear,serialise,120,120,120,59,59,46,42,35,23,23,23,59
linear,sieve,85,85,85,85,85,82,84,81,68,68,68,82
linear,simple_analyzer,TO/OOM,TO,TO,TO,TO,TO,TO,TO,TO,TO,TO,TO
linear,tak,104,104,104,104,104,94,89,84,66,66,66,104
linear,times10,101,101,101,101,101,101,97,97,80,80,80,101
linear,unify,1601,1395,1395,1414,1414,784,TO,TO,519,TO,519,1361
linear,zebra,OOM,TO,TO,4261,4261,TO,TO,TO,TO,TO,TO,TO
\end{filecontents*}
\csvreader[
  respect all,
  tabular=ll*{12}{c},
  table head={
    \toprule
    property & program &


    shlin2 & shlin2 & shlin2 &
    shlin & shlin & shlin &
    shlin &  shlin &
    sharing & sharing & 
    share & shfrlin \\

    &&
    (+) & +M & M &
    ++ & + &  &
    +M &  M &
    (+) & (+)M & 
     &  \\

    \midrule
  },
  late after line=\\,
  table foot=\bottomrule
]{report_precision_best_5.csv}{
  property=\prop,
  program=\prog,
  as_shlin2(+)=\cA,
  as_shlin2_opt_mgu=\cB,
  as_shlin2_noopt_mgu=\cD,
  as_shlin_opt_opt=\cE,
  as_shlin_opt=\cF,
  as_shlin_noopt=\cH,
  as_shlin_opt_mgu=\cI,
  as_shlin_noopt_mgu=\cK,
  as_sharing(+)=\cL,
  as_sharing_mgu(+)=\cN,
  share=\cP,
  shfrlin=\cQ
}{%
  \prop & \prog &
  \cA & \cB & \cD &
  \cE & \cF & \cH &
  \cI & \cK &
  \cL & \cN & 
  \cP & \cQ
}
\end{adjustbox}
\caption{\label{table:precision}\small Precision results for all analyzed programs and configurations. 
+ indicates the use of optimal mgu, while (+) indicates identical results for standard and optimal mgu (which are therefore collapsed in a single column); M denotes the use of matching instead of mgu; ++ denotes the use of both optimal mgu and optimal matching. 
\emph{OOM} denotes an out-of-memory condition, while \emph{TO} denotes a timeout.}
\end{table}

\section{Conclusions}
\label{sec:conclusions}
The results of our experimental evaluation can be summarized as follows.
\begin{description}
    \item[Optimality.] The main outcome is that the use of optimal abstract operators does not compromise the feasibility of the analysis. Even if their definitions are more convoluted, optimal abstract operators 
    can be implemented efficiently in practice. In many cases, they even lead to faster analyses. This behavior is likely a peculiarity of \emph{may} abstract domains: the additional precision provided by optimal operators often results in smaller abstract objects. As a consequence, the effort required to compute the optimal operators is compensated by the reduced size of the resulting abstract objects. On the other hand, the benefits of optimal operators are not uniform and tend to be significant only in selected cases. 
    \item[Matching.] The use of matching consistently outperforms unification. It improves both precision and performance, and in several cases enables analyses that would otherwise time out when relying solely on unification. As a result, matching proves to be a particularly effective choice in practice.
\end{description}
As future work, we plan to extend the experimental evaluation by considering additional benchmark programs and improving the implementation of some abstract operators, in particular the optimal mgu for $\ShLin$ and the matching operator for \KingShLin, which still follow an unoptimized \emph{generate-and-test} approach.
We also plan to investigate the implementation in \CiaoPP of numerical analyses, such as the parallelotope-based analysis by \cite{AmatoRS17-scp}, and more advanced fixpoint solvers, such as the ones implemented in ScalaFix \citep{AmatoS23-fm}.



%

\section*{Acknowledgments}
We thank Manuel Hermenegildo and José F. Morales for valuable discussions on the inner workings of \CiaoPP.

\section*{Competing interests}
The authors declare none.

\bibliographystyle{tlplike2}
\bibliography{asbiblio2_local}

\end{document}